\documentclass[twocolumn]{aastex62}
\usepackage{CJK}

\newcommand\aastex{AAS\TeX}

\submitjournal{ApJ}

\shorttitle{\aastex\ CO multi-line observations of G240.31+0.07}
\shortauthors{Liu et al.}

\begin{document}
\begin{CJK*}{UTF8}{gbsn}
\title{An Isothermal Outflow in High-mass Star-forming Region G240.31+0.07}

\correspondingauthor{Keping Qiu}
\email{kpqiu@nju.edu.cn}

\author[0000-0002-4774-2998]{Junhao Liu (刘峻豪)}
\affil{School of Astronomy and Space Science, Nanjing University, 163 Xianlin Avenue, Nanjing 210023, P.R.China}
\affil{Key Laboratory of Modern Astronomy and Astrophysics (Nanjing University), Ministry of Education, Nanjing 210023, P.R.China}

\author[0000-0002-5093-5088]{Keping Qiu}
\affil{School of Astronomy and Space Science, Nanjing University, 163 Xianlin Avenue, Nanjing 210023, P.R.China}
\affil{Key Laboratory of Modern Astronomy and Astrophysics (Nanjing University), Ministry of Education, Nanjing 210023, P.R.China}

\author{Friedrich Wyrowski}
\affiliation{Max-Planck-Institut f{\"u}r Radioastronomie, Auf dem H{\"u}gel 69, 53121 Bonn, Germany}

\author{Karl Menten}
\affiliation{Max-Planck-Institut f{\"u}r Radioastronomie, Auf dem H{\"u}gel 69, 53121 Bonn, Germany}

\author{Rolf G{\"u}sten}
\affiliation{Max-Planck-Institut f{\"u}r Radioastronomie, Auf dem H{\"u}gel 69, 53121 Bonn, Germany}

\author[0000-0002-6368-7570]{Yue Cao}
\affil{School of Astronomy and Space Science, Nanjing University, 163 Xianlin Avenue, Nanjing 210023, P.R.China}
\affil{Key Laboratory of Modern Astronomy and Astrophysics (Nanjing University), Ministry of Education, Nanjing 210023, P.R.China}

\author[0000-0001-7687-9320]{Yuwei Wang}
\affil{School of Astronomy and Space Science, Nanjing University, 163 Xianlin Avenue, Nanjing 210023, P.R.China}
\affil{Key Laboratory of Modern Astronomy and Astrophysics (Nanjing University), Ministry of Education, Nanjing 210023, P.R.China}

\begin{abstract}
We present Atacama Pathfinder EXperiment (APEX) observations toward the massive star-forming region \objectname{G240.31+0.07} in the CO J = 3--2, 6--5, and 7--6 lines. We detect a parsec-sized, bipolar, and high velocity outflow in all the lines, which allow us, in combination with the existing CO J = 2--1 data, to perform a multi-line analysis of physical conditions of the outflowing gas. The CO 7--6/6--5, 6--5/3--2, and 6--5/2--1 ratios are found to be nearly constant over a velocity range of $\sim$5--25 km s$^{-1}$ for both blueshifted and redshifted lobes. We carry out rotation diagram and large velocity gradient (LVG) calculations of the four lines, and find that the outflow is approximately isothermal with a gas temperature of $\sim$50 K, and that the the CO column density clearly decreases with the outflow velocity. If the CO abundance and the velocity gradient do not vary much, the decreasing CO column density indicates a decline in the outflow gas density with velocity. By comparing with theoretical models of outflow driving mechanisms, our observations and calculations suggest that the massive outflow in G240.31+0.07 is being driven by a wide-angle wind and further support a disk mediated accretion at play for the formation of the central high-mass star.

\end{abstract}


\section{Introduction}
Molecular outflows, mostly observed in rotational transitions of CO, are a common phenomenon associated with young stellar objects (YSOs) of all the masses \citep{2001ApJ...552L.167Z, 2002A&A...383..892B, 2004A&A...426..503W, 2005AJ....129..330W,  2015MNRAS.453..645M}. Although molecular outflows have been extensively observed, their driving mechanism is still not well understood \citep{2007prpl.conf..245A,2014prpl.conf..451F,Bally16}. 

Among various models developed to explain molecular outflows in low-mass YSOs,  those interpret outflows as ambient gas being swept up by an underlying wide-angle wind \citep{1991ApJ...370L..31S, 2000prpl.conf..789S}, or by a collimated jet through bow shocks \citep{1993A&A...278..267R, 1993ApJ...414..230M}, have attracted most interests \citep[e.g.,][and references therein]{2001ApJ...557..429L}. While the wind-driven and jet-driven models can both explain some of observed features, none of them is able to explain a full range of morphologies and kinematics of different types of outflows \citep{2000ApJ...542..925L, 2002ApJ...576..294L}. This has motivated a number of theoretical efforts dedicated to unifying a collimated jet and a wide-angle wind in understanding the driving mechanism of low-mass outflows \citep{2006ApJ...641..949B, 2006MNRAS.365.1131P, 2006ApJ...649..845S, 2007prpl.conf..277P, 2008ApJ...676.1088M}.

Massive molecular outflows, which are associated with high-mass YSOs, are even less clear than their low-mass counterparts. Due to their typically large distances, they are far more difficult to resolve. Nevertheless, some observations appear to suggest that massive outflows have morphologies and kinematics similar to those of low-mass outflows \citep[e.g.,][]{1998ApJ...507..861S, 2002A&A...387..931B, 2009ApJ...696...66Q, 2011MNRAS.415L..49R}. Theoretically, there is few work capable of modeling massive outflows to time and size scales ready to be compared with observations. Consequently, several key questions, e.g., how massive outflows are driven, and how they intrinsically differ from their low-mass counterparts, are yet to be addressed. 

Previous observational studies of molecular outflows are mostly based on low-J rotational transitions of CO ($J_{\mathrm{up}}\leq3$, with upper-state energies, $E_{\mathrm{up}}$, $\lesssim30$~K), which are readily excited at low temperatures and can be easily observed by ground-based facilities, to characterize the morphology and kinematics of the relatively cold and extended molecular gas. Due to the Earth's atmosphere, mid-J CO lines (referring to CO J = 6--5 and 7--6 throughout this paper, with  $E_{\mathrm{up}}$ up to 150 K), which are less affected by the ambient gas, are not commonly observed. In several studies, mid-J CO transitions have been reported to trace a warm gas ($>$50 K) in outflows of low-mass and intermediate-mass YSOs \citep{2009A&A...501..633V, 2009A&A...507.1425V, 2012A&A...542A..86Y, 2016A&A...587A..17V}. By comparing CO multi-line observations (both low-J and mid-J) with the results of radiative transfer models, the physical properties (temperature, density, and CO column density) of the outflowing gas could be better constrained \citep{2015A&A...581A...4L}. 

\objectname{G240.31+0.07} (hereafter \objectname{G240}) is an active high-mass star-forming region with a far-infrared luminosity of 10$^{4.6}~L_\sun$ at a distance of 5.4~kpc \citep{1998AJ....116.1897M,2014ApJ...790...99C, 2015PASJ...67...69S}. It harbors at least one ultracompact H{\scriptsize II} region and is associated with OH and H$_2$O masers \citep{1993AJ....105.1495H,1997MNRAS.289..203C,1998AJ....116.1897M,1999ApJS..123..487M,2003MNRAS.341..551C,2011AJ....142..147T}. The cloud velocity with respect to the local standard of rest, $v_{\mathrm{cloud}}$, was found to be 67.5 km\,s$^{-1}$ from C$^{18}$O J = 2--1 observations \citep{2003A&A...412..175K}. \citet{2009ApJ...696...66Q} presented a detailed high-resolution study of CO and $^{13}$CO J = 2--1 emissions in \objectname{G240}, and detected a bipolar, wide-angle, and quasi-parabolic molecular outflow with a gas mass of order $100~M_{\odot}$. \citet{2013A&A...559A..23L} theoretically investigated the possibility that the \objectname{G240} outflow results from interaction between a wide-angle wind and the ambient gas, and suggested that turbulent motion helps the wind to entrain the gas into the outflow. More recently, \citet{2014ApJ...794L..18Q} reported the detection of an hourglass magnetic field aligned within 20${\degr}$ of the outflow axis.

\section{Observations}

The APEX CO J = 6--5 and 7--6 observations were performed on 2010 July 11 with the Carbon Heterodyne Array of the MPIfR (CHAMP$^+$), which is a dual-frequency heterodyne array consisting of $2\times7$ pixels for spectroscopy in the 690~GHz and 810~GHz atmospheric windows \citep{2006SPIE.6275E..0NK}. The APEX beams at these two frequencies are about $9.\!''0$ and $7.\!''7$. We tuned the receiver array to simultaneously observe CO 6--5 at 691.4731 GHz and CO 7--6 at 806.6518 GHz. Signals from each CHAMP$^+$ pixel were processed by two overlapping Fast Fourier Transform (FFT) spectrometers, which were configured to provide a spectral resolution of 0.73~MHz and a bandwidth of 2.4~GHz. The observations were made in the on-the-fly (OTF) mode. We obtained a $81''\times39''$ map with a cell size of $3''$ for each line, and the long axis of the map was titled by $36^{\circ}$ west of north to roughly follow the outflow axis orientation \citep{2009ApJ...696...66Q}. The intensity scale was converted from the calibrated antenna temperature, $T_{\rm A}^*$, to the main beam antenna temperature, $T_{\rm mb}$, with beam efficiencies of 0.41 at 690~GHz and 0.40 at 810~GHz, measured from observations of planets. We smoothed the data into 2~km\,s$^{-1}$ velocity channels, resulting in root-mean-square (RMS) noise levels of about 0.2~K in $T_{\rm mb}$ for CO 6--5 and 0.5~K for CO 7--6.

The APEX CO J = 3--2 observations were made on 2011 October 16 using the First Light APEX Submillimeter Heterodyne (FLASH), which is a dual-frequency receiver operating simultaneously in the 345~GHz and 460~GHz atmospheric windows \citep{2006A&A...454L..21H}. The APEX beams at these two frequencies are about $17.\!''7$ and $13.\!''6$. We tuned the receiver to observe CO 3--2 at 345.7960 GHz and CO 4--3 at 461.0408 GHz. The CO 4--3 data were not useful due to the limited weather conditions, and will not be further discussed in this work. FLASH outputs signals to FFT spectrometers which provided a spectral resolution of 76~kHz and a bandwidth of 4~GHz. We obtained a $96''\times78''$ OTF map with a cell size of $6''$. The long axis of the map was also titled by $36^{\circ}$ west of north. The intensity scale in $T_{\rm mb}$ was derived with a beam efficiency of 0.65 at 345~GHz. Again we smoothed the data into 2~km\,s$^{-1}$ velocity channels, and the corresponding RMS noise level is 0.04~K in $T_{\rm mb}$.

 The APEX observations are jointly analyzed with existing high resolution CO 2--1 data, and the latter have an RMS noise level of 0.08~K for a spectral resolution of 2~km\,s$^{-1}$ \citep{2009ApJ...696...66Q}. We summarize in Table \ref{tab:lines} some of the observational parameters for the CO lines used in the following analyses.

\begin{deluxetable}{ccCccc}[t!]
\tablecaption{Observations \label{tab:lines}}
\tablecolumns{6}
\tablewidth{0pt}
\tablehead{
\colhead{Transition} &
\colhead{Frequency} &
\colhead{Beam size} & 
\colhead{Cal. unc. \tablenotemark{a}} & 
\colhead{$\sigma_{\rm rms}$\tablenotemark{b}} & 
\colhead{$\eta_{\rm mb}$\tablenotemark{c}} \\
\colhead{} & \colhead{GHz} &
\colhead{$\arcsec$} & \colhead{\%} & \colhead{K} & \colhead{}
}
\startdata
2--1 & 230.5380 & 3.93 $\times$ 3.10 & 10 & 0.08& $-$\\
3--2 & 345.7960 & 17.7 & 15 & 0.04 & 0.65 \\
6--5 & 691.4731 & 9.0 & 20 & 0.20 & 0.41 \\
7--6 & 806.6518 & 7.7 & 30 & 0.50 & 0.40 \\
\enddata
\tablenotetext{a}{Adopted flux calibration uncertainties.}
\tablenotetext{b}{RMS noise leves for 2 km\,s$^{-1}$ velocity channels.}
\tablenotetext{c}{Beam efficiency.}
\end{deluxetable}

\section{Results}
\subsection{CO emission maps}

\begin{figure*}[htbp]
\includegraphics[scale=.65]{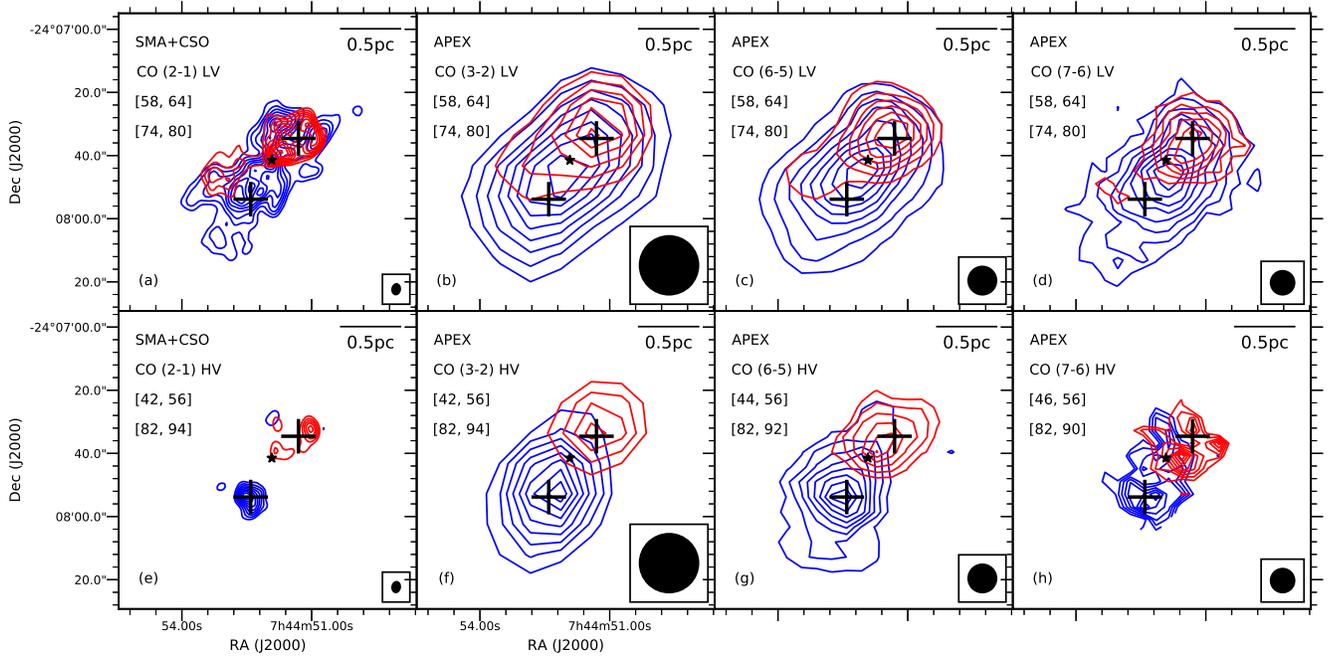}
\caption{(a)-(d) Low-velocity CO J = 2--1, 3--2, 6--5, and 7--6 emissions, integrated from 58 to 64 km s$^{-1} $ for the blueshifted lobe (blue) and from 74 to 80 km s$^{-1}$ for the redshifted lobe (red); (e)-(f) High-velocity CO J = 2--1 and 3--2 emissions,  integrated from 42 to 56 km s$^{-1} $ for the blueshifted lobe (blue) and from 82 to 94 km s$^{-1}$ for the redshifted lobe (red); (g) High-velocity CO J = 6--5 emission, integrated from 44 to 56 km s$^{-1} $ for the blueshifted lobe (blue) and from 82 to 92 km s$^{-1}$ for the redshifted lobe (red) (h) High-velocity CO J = 7--6 emission, integrated from 46 to 56 km s$^{-1} $ for the blueshifted lobe (blue) and from 82 to 90 km s$^{-1}$ for the redshifted lobe (red). For (a)-(g), the contour levels start from 20\% and continue at steps of 10\% of the peak emission. For (h), the contour levels start from 30\% and continue at steps of 10\% of the peak emission. Edge channels are masked out because of high noise levels. The black star marks the position of a H$_2$O maser spot which is associated with IRAS 07427-2400 \citep{2015PASJ...67...69S}. The beam of each observational dataset is shown in the lower right corner of each panel. \label{fig:figcontour}}
\end{figure*}

The CO J = 3--2, 6--5, and 7--6 emissions are detected (with obvious outflow signatures and peak intensities $>2\sigma_{\rm rms}$) in velocity ranges from 42 to 94 km s$^{-1}$, 44 to 92 km s$^{-1}$, and 46 to 90 km s$^{-1}$, respectively. Figure \ref{fig:figcontour} shows the integrated low-velocity (LV) and high-velocity (HV) emissions of the four lines. The velocity ranges chosen to highlight the LV and HV components of the outflowing gas follow those in \citet{2009ApJ...696...66Q}, except that the channels with no detections were excluded for the HV component. The morphologies of the bipolar outflow seen in the CO J = 3--2, 6--5, and 7--6 lines are very similar. Due to the coarser angular resolutions, the wide-angle structure seen in the high-resolution CO J = 2--1 image is not resolved in the CO 3--2, 6--5, and 7--6 maps.

\subsection{Physical conditions of the outflow}

\begin{figure}[tbp]
\plotone{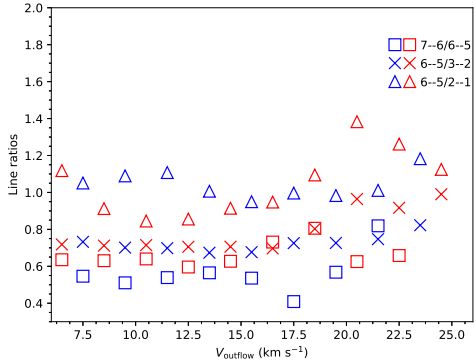}
\caption{Ratios of the main-beam temperatures of different CO lines at different velocities. Blue symbols denote the measurements from the blueshifted lobe, and red symbols the redshifted lobe. The $V_{\mathrm{outflow}}$ shown here is related to the cloud velocity $v_{\mathrm{cloud}}$ by the relation: $V_{\mathrm{outflow}}=|v_{\mathrm{outflow}} - v_{\mathrm{cloud}}|$, where $v_{\mathrm{outflow}}$ is the outflow velocity with respect to the local standard of rest. \label{fig:figratio}}
\end{figure}

The physical conditions of the outflow can be constrained by comparing the observed line intensities with the results of radiative transfer calculations. To perform such analyses, the CO J = 2--1, 6--5, and 7--6 maps are convolved to the beam of the CO 3--2 map. The RMS noise levels are $\sim$0.004 K, $\sim$0.04 K and $\sim$0.1 K for the convolved CO J = 2--1, 6--5 and 7--6 data, respectively. For both lobes of the outflow, the CO line intensities are measured at approximately the peak positions of the HV components of the convolved CO maps (marked as two crosses in each panel of Figure \ref{fig:figcontour}), and are then used in the following analyses. Figure \ref{fig:figratio} shows the ratios of $T_{\mathrm{mb}}$ for different CO lines as functions of velocity. The CO 7--6/6--5, 6--5/3--2, and 6--5/2--1 ratios are almost constant over a velocity range of $\sim$5--25 km s$^{-1}$ with respect to the cloud velocity, with ratio variations $<30\%$. The average values of the 7--6/6--5, 6--5/3--2, and 6--5/2--1 ratios are $\sim$0.6, $\sim$0.7, and $\sim$1.0, respectively. We only use channels of $\le$60 km s$^{-1}$ and $\ge$74 km s$^{-1}$ to avoid contaminations from the ambient gas, and exclude channels of $<$46 km s$^{-1}$ and $>$90 km s$^{-1}$ because of their low signal-to-noise ratios. Since we do not correct the observed line intensities for unknown beam filling factors, the derived CO column densitiy ($N$), gas temperature ($T$), and gas density ($n$) should be considered as beam-averaged values.

\begin{figure}[tbp]
\plotone{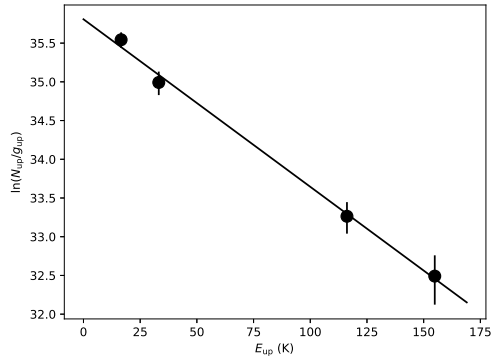}
\caption{A rotation diagram for CO at 84 km s$^{-1}$. The fitted line shows the Boltzmann distribution of the rotational populations. The line represents a rotational temperature of 48.5 K and a total column density of 2.0 $\times$ 10$^{15}$ cm$^{-2}$. The black solid circles show the data with error bars. \label{fig:figrd}}
\end{figure}

We first perform a simple RD analysis \citep{1999ApJ...517..209G} to estimate the temperature and CO column density of the outflowing gas under the assumption of local thermal equilibrium (LTE). Considering that the $^{13}$CO J = 2--1 emission was only marginally detected or not detected at our considered velocities \citep{2009ApJ...696...66Q}, we assume that the CO emissions are optically thin. The population of each energy level is given by 
\begin{equation}
N_{\mathrm{up}} = \frac{N_\mathrm{CO}}{Z} g_\mathrm{up} e^{-E_\mathrm{up}/kT_\mathrm{kin}},
\end{equation}
where $N_\mathrm{up}$ is the column density in the upper state, $g_\mathrm{up}$ the statistical weight of the upper state, $E_\mathrm{up}$ the upper energy level, $k$ the Boltzmann constant, and $Z$ is the partition function. The rotation diagram for CO at 84 km s$^{-1}$ is shown in Figure \ref{fig:figrd} as an example. An LTE model at 48.5 K could account for the measurements from the four lines. Other velocity channels show similar rotation diagrams. The error bars in Figure \ref{fig:figrd} take into account both the flux calibration uncertainties and the RMS noises. 

\begin{figure*}[!tbp]
\gridline{\fig{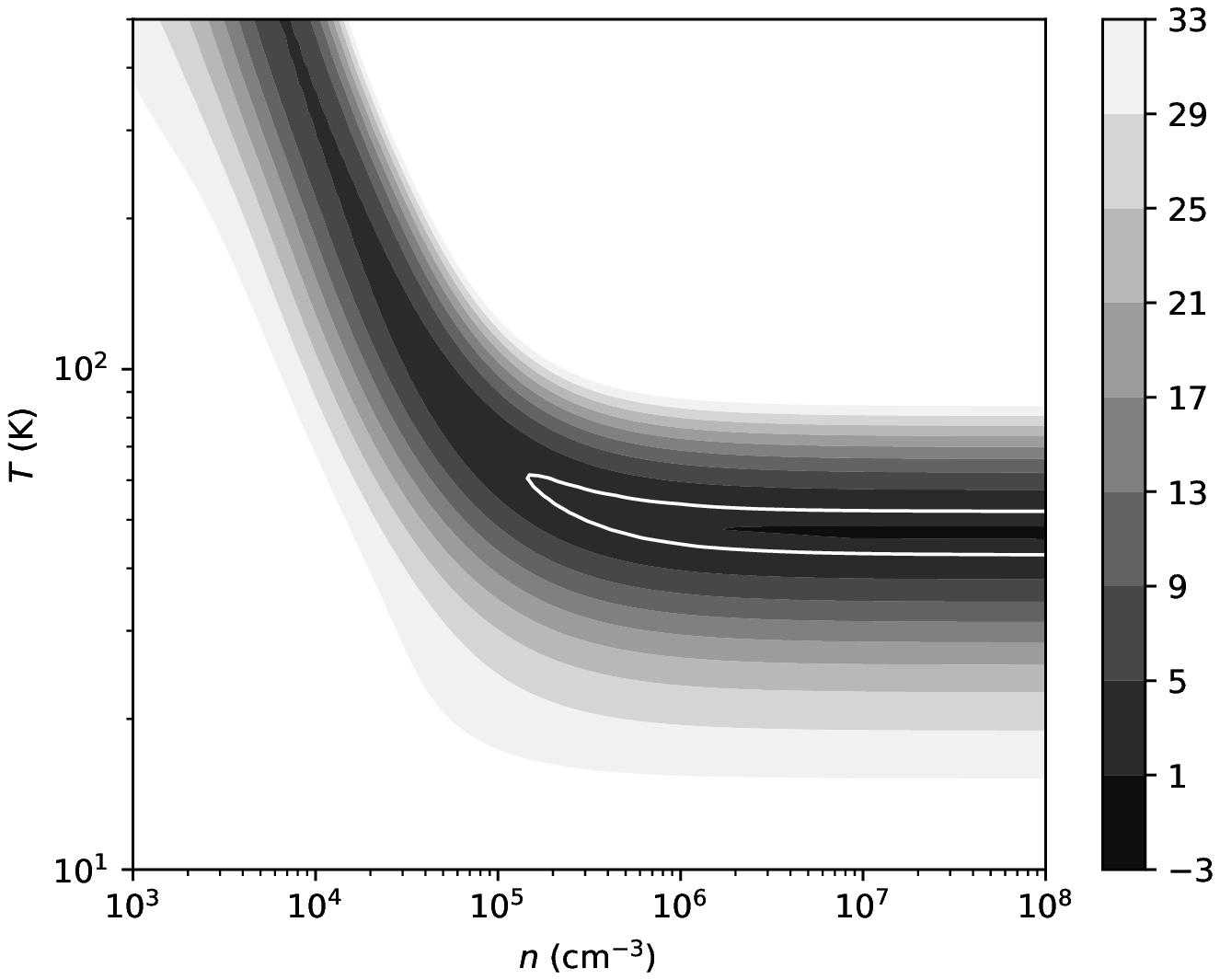}{0.5\textwidth}{(a)}
        \fig{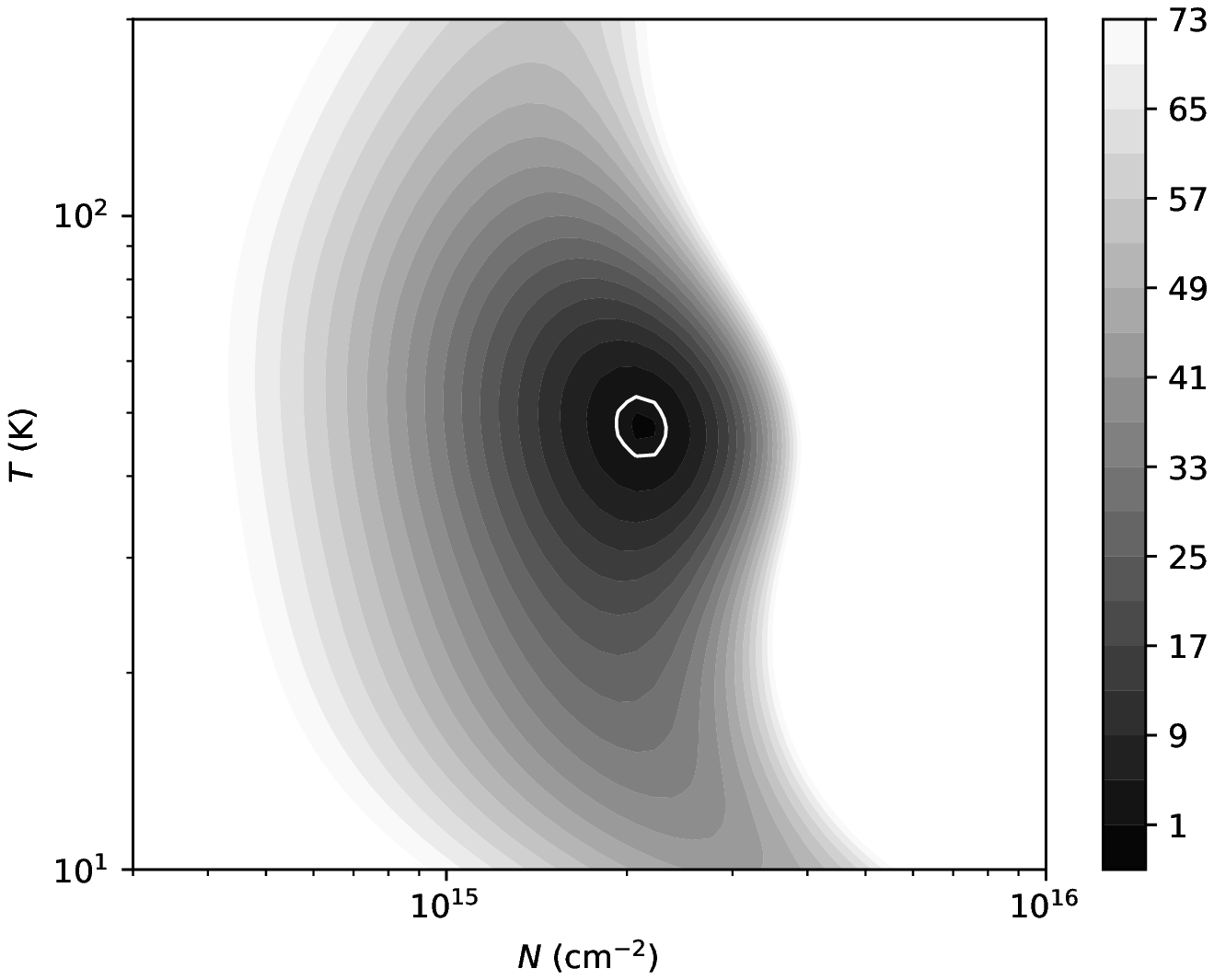}{0.5\textwidth}{(b)}}
 \gridline{\fig{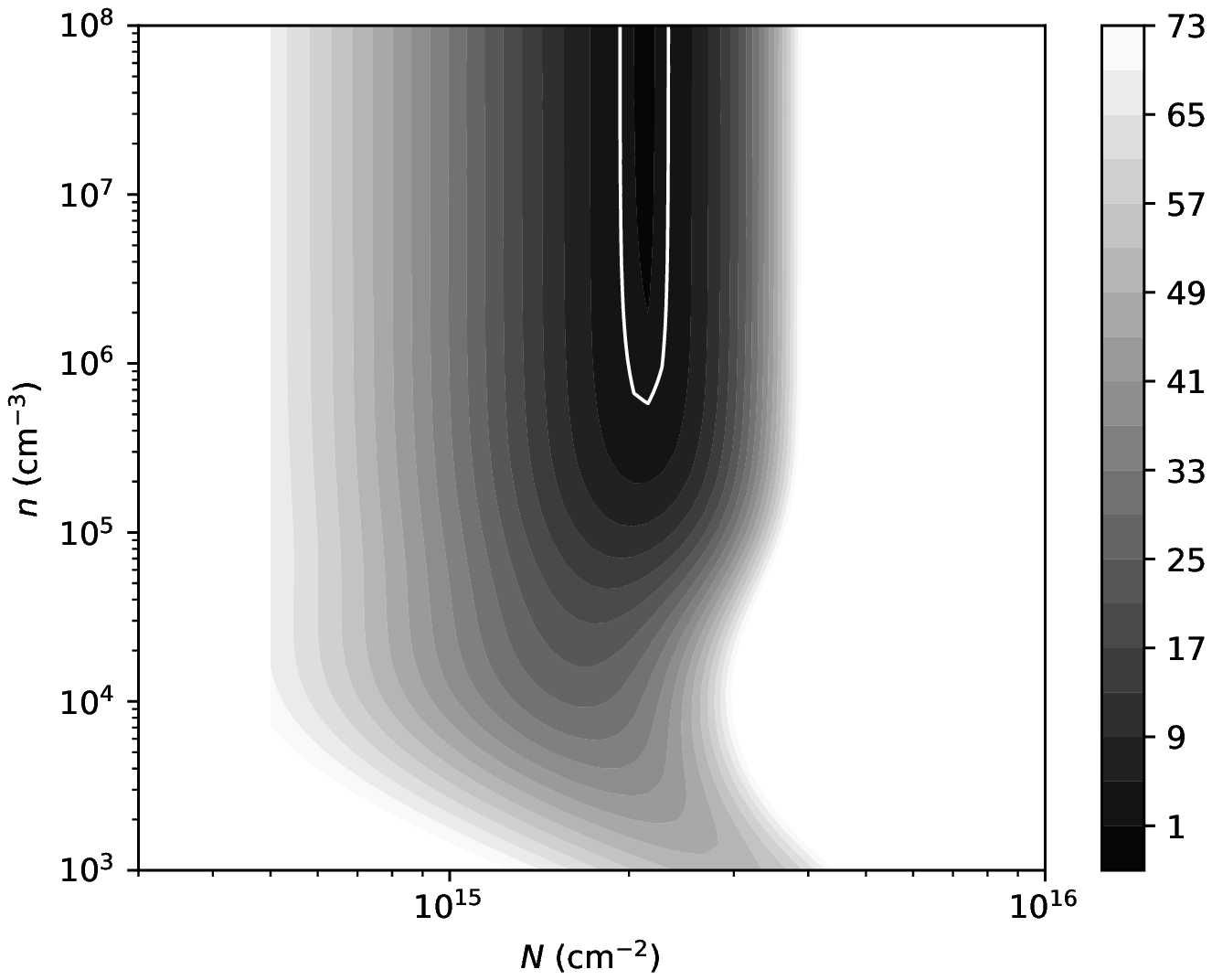}{0.5\textwidth}{(c)}
      }
\caption{(a)-(c) The $\chi^2$ distribution at 84 km s$^{-1}$ in the [$T$, $n$], [$T$, $N$] and [$n$, $N$] planes, with the third parameter fixed to the value of the best fitting result at this velocity. The lower limit of gas density is 1.8 $\times$ 10$^{5}$ cm$^{-3}$. The best-fit solution is obtained for $T$ =  46.1 K and $N$ = 2.2 $\times$ 10$^{15}$ cm$^{-2}$. The $\chi^2_{\mathrm{red}}$ of the best-fit solution is 1.00. The Solid white contours show the 1$\sigma$ confidence levels. \label{fig:figchi}}
\end{figure*}

To better constrain the gas temperature, density, and the CO column density without assuming LTE and optically thin emissions, we then carry out radiative transfer calculations of the four lines using the RADEX code, which adopts the LVG approximation \citep{2007A&A...468..627V}. We build a large grid of LVG models by varying the three parameters ($n$, $T$, and $N$), and obtain the best fitting results by $\chi^2$ minimization in comparing the observations with the models. With four lines observed and three parameters to constrain, the fitting has one degree of freedom. In Figure \ref{fig:figchi}, the fitting results at 84 km s$^{-1}$ are shown as an example. Figure \ref{fig:figchi}(a) and Figure \ref{fig:figchi}(b) show that the gas temperature is well constrained to $\sim$ 50 K. In Figures \ref{fig:figchi}(b) and \ref{fig:figchi}(c), the CO column density is stringently constrained to 2.2 $\times$ 10$^{15}$ cm$^{-2}$. The $\chi^2$ distribution in Figure \ref{fig:figchi}(a) and \ref{fig:figchi}(c) indicate that in the best LVG models, the gas density is high enough to thermalize the emissions, and thus only lower limits of the gas density could be derived from the $\chi^2$ minimization. This implies that the LTE assumption adopted by the above RD analysis is valid, and hence we obtain similar gas temperatures and CO column densities from the RD and LVG analyses. Similar $\chi^2$ distribution patterns are found at other velocities. The reduced $\chi^2$ ($\chi^2_{\mathrm{red}}$) of the best fitting results varies from 0.10 to 1.72 at different velocities. The representative uncertainty of each parameter is derived from the 1$\sigma$ confidence region in the 3D parameter space at the velocities where $\chi^2_{\mathrm{red}} \sim 1$: the relative uncertainty of the CO column density is $\sim$10\%; the temperature uncertainty is $\sim$10 K; and the gas density is $>$10$^5$ cm$^{-3}$ over the entire velocity range. The modeling results also predict that the four transitions are optically thin in the outflowing gas. Figure \ref{fig:figsed} shows a comparison of the observed CO intensities with the LVG modeling results in each velocity bin. The best fitting solution agrees well with the observations at all the velocities.

\begin{figure*}[!tbp]
\gridline{\fig{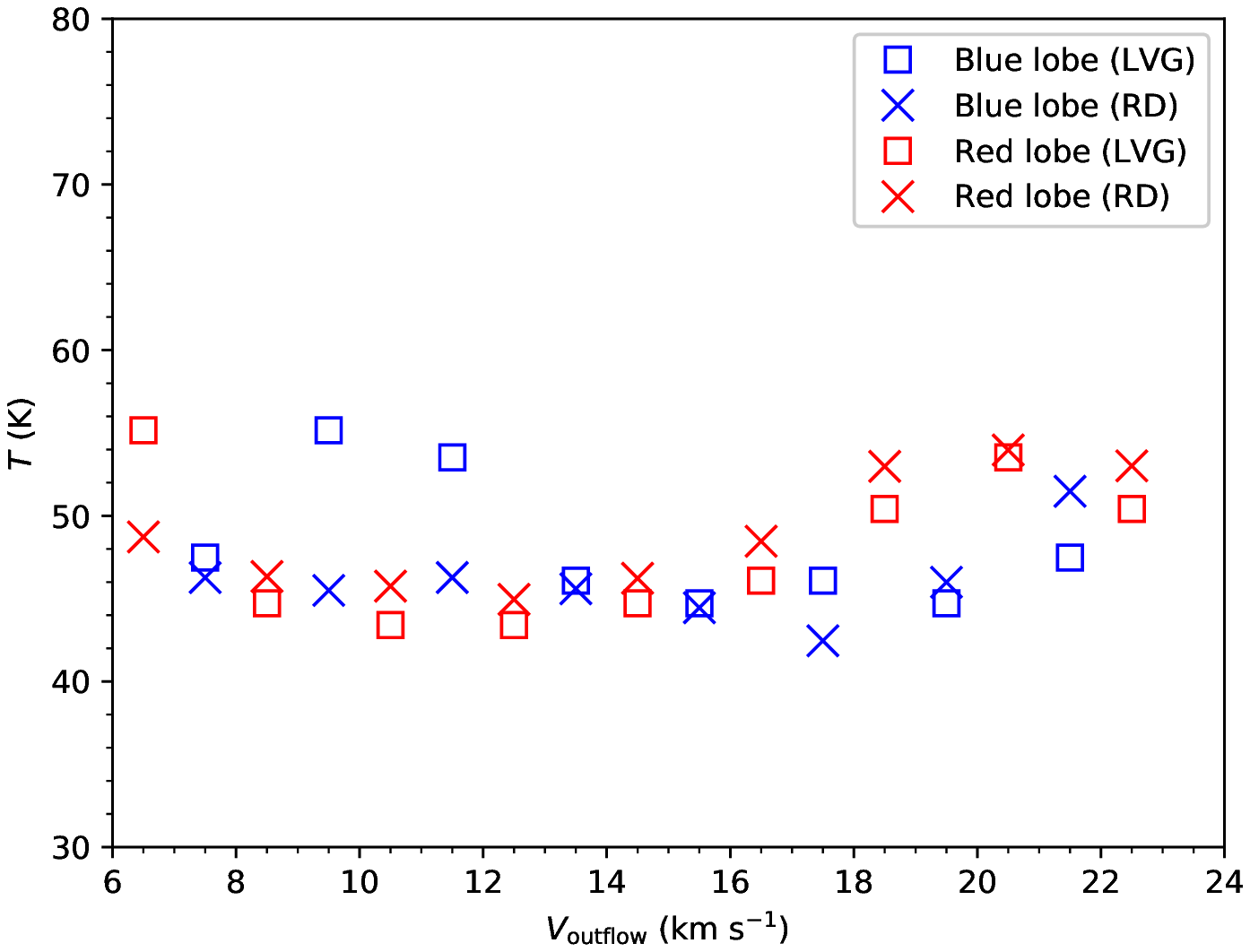}{0.5\textwidth}{(a)}
          \fig{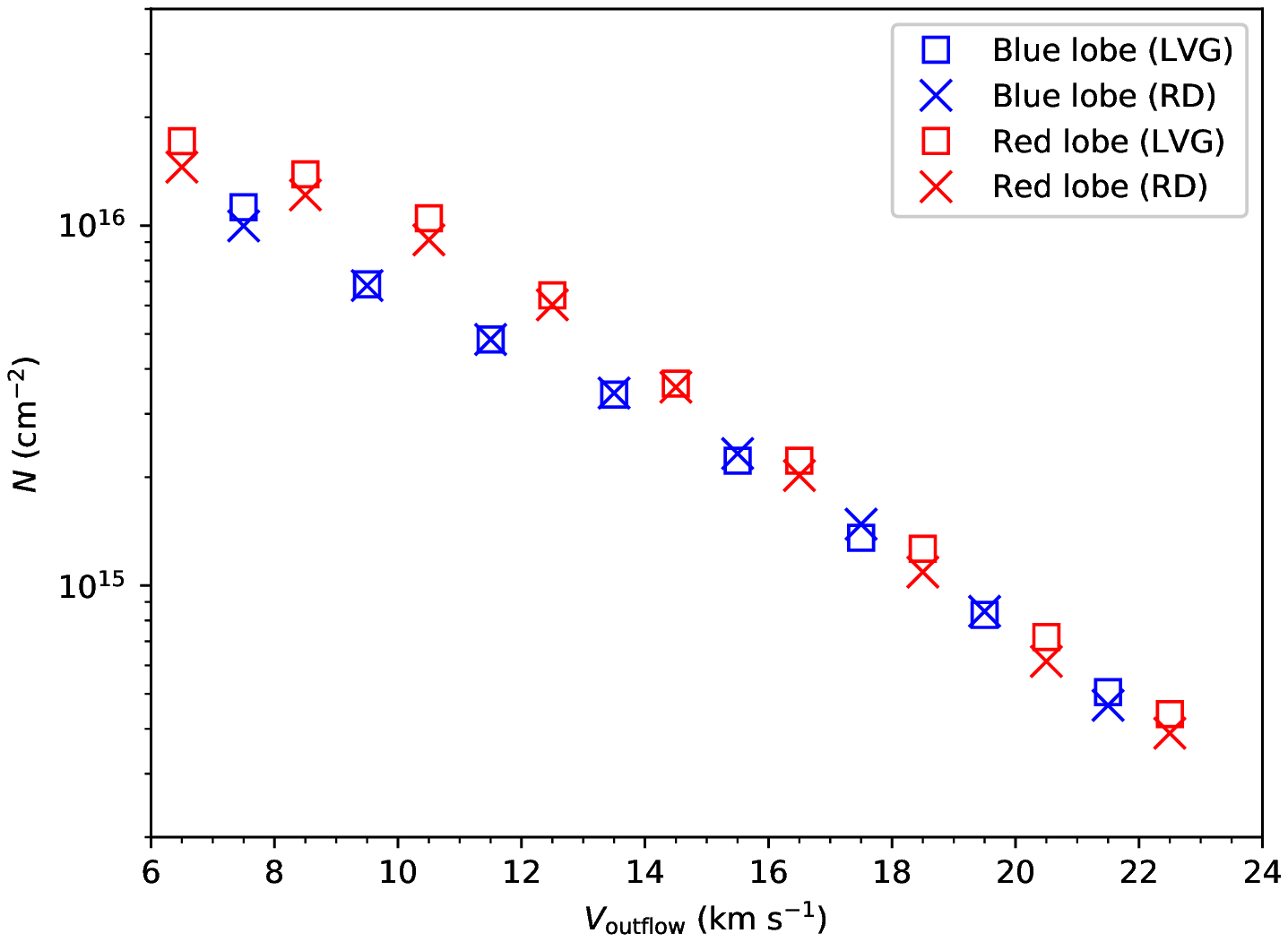}{0.5\textwidth}{(b)}
          }
\caption{$T$-$V$ and $N$-$V$ diagrams of the G240 outflow, estimated from the RD analysis (blue ``x'' markers for the blue lobe and red ``x'' markers for the red lobe) and the LVG analysis (blue open squares for the blue lobe and red open squares for the red lobe). \label{fig:figrelation}}
\end{figure*}

The variation of the physical conditions of the outflowing gas as a function of velocity is of great interests to our understanding of the driving mechanism of the outflow. With the RD and LVG analyses we have estimated the gas temperature, density, and CO column density at different velocities. Figure \ref{fig:figrelation} shows the temperature-velocity ($T$-$V$) and CO column density-velocity ($N$-$V$) relations. In Figure \ref{fig:figrelation}(a), the $T$-$V$ diagram shows that the emission of the outflow is well accounted for by gas with a temperature of $\sim$50 K; the apparent variation of $<$10 K is within the uncertainties, indicating that the outflowing gas is approximately isothermal. Figure \ref{fig:figrelation}(b) shows that the CO column density clearly decreases from $2\times 0^{16}$ cm$^{-2}$ to $4\times10^{14}$ cm$^{-2}$ as the outflow velocity increases from 7 km s$^{-1}$ to 22 km s$^{-1}$. We have made RD and LVG calculations of the CO lines in different positions, and obtain similar $T$-$V$ and $N$-$V$ relations. Thus, it appease that the $T$-$V$ and $N$-$V$ trends are independent of positions within the outflow region.

\section{Discussion}\label{discussion}
The morphology and kinematics of molecular outflows have been widely studied based on single spectral line observations. However, the physical conditions of the outflow gas are still not well constrained. In paticular, how the physical conditions vary with the outflow velocity is poorly understood. To precisely constrain the physical properties of molecular outflows, CO observations across a wide range of energy levels are needed. Previous studies show that CO transitions of $J_{\mathrm{up}} > 9$ (high-J) and with $J_{\mathrm{up}} \le 9$ should be fitted with different gas components, and that the hotter, denser gas component traced by high-J CO lines only constitutes a small amount of the gas in the molecular outflow \citep{2013A&A...555A...8G, 2015A&A...581A...4L}. Thus, the driven mechanism of the bulk of the molecular outflow is best studied in lines of $J_{\mathrm{up}} \le 9$. Our multi-line analysis with both low-J and mid-J CO line observations have constrained the CO column density and the temperature of the G240 outflow as functions of the outflow velocity, showing a clear example on the physical conditions of a representative well-defined bipolar wide-angle molecular outflow in a $>$10$^4~L_\sun$ massive star-forming region \citep{2009ApJ...696...66Q}.

\subsection{Temperature-velocity relation}
The variation of the outflow temperature with the outflow velocity have been studied only in a handful of sources. \citet{2009A&A...501..633V} and \citet{2012A&A...542A..86Y} performed LVG calculations on single-dish CO J = 3--2 and 6--5 lines, and suggested that there is little or no temperature change within outflow velocity ranges of $<$10--15 km s$^{-1}$ in the outflow associated with low-mass protostars HH46 IRS1 and NGC 1333 IRAS 4A/4B. Considering that \citet{2009A&A...501..633V} have observed rising CO 3--2/6--5 ratios at more extreme velocities, and that the line wing ratios of CO 3--2/6--5 observed by \citet{2012A&A...542A..86Y} have relatively large variations, their conclusions of ``constant'' outflow temperatures are not robust. The temperature of the outflowing gas as a function of velocity in the outflow associated with a high-mass star-forming region was first derived by \citet{2012ApJ...744L..26S}, who performed an LVG analysis on interferometer CO J = 2--1 and 3--2 observations of the extremely high velocity outflow associated with the high-mass star-forming region G5.89-0.39, and found an increasing temperature with outflow velocity for outflow velocities up to 160 km s$^{-1}$. Based on the variation of the CO 2--1/1--0 line ratio at a resolution of 32.5$\arcsec$ and assuming LTE, \citet{2018RAA....18...19X} found that the excitation temperature of the outflow in the massive star-forming region IRAS 22506+5944 increases from low ($\sim$5 km s$^{-1}$) to moderate ($\sim$8-12 km s$^{-1}$) velocities, and then decrease at higher velocities ($<$30 km s$^{-1}$). However, \citet{2012ApJ...744L..26S} and \citet{2018RAA....18...19X} have only used low-J CO lines with small energy ranges, which are not sufficient to trace the relatively warm and dense gas. Moreover, all of the above mentioned works have made assumptions about other parameters (e.g., gas density, canonical CO fractional abundance, or velocity gradient) or the equilibrium state (e.g., LTE) of the outflow to infer the $T$-$V$ relations from only two CO lines, while these assumptions might not be necessarily valid. Recently, the outflow properties were more precisely determined via multi-line CO studies by \citet{2015A&A...581A...4L}, who performed an LVG analysis on the outflow cavity ($<$40 km s$^{-1}$) of the intermediate-mass Class 0 protostar Cepheus E-mm and revealed that the outflowing gas traced by CO transitions of $J_{up} \le 9$ is nearly isothermal. Our analysis, for the first time, reveals the temperature-velocity relation in a high-mass star-forming region with both low-J and mid-J observations via the LVG calculation. The results show that the G240 outflow is approximately isothermal with a gas temperature of $\sim$50 K within an outflow velocity range of $<$23 km s$^{-1}$. This isothermal state is similar to the behavior of the outflowing gas (traced by CO transitions of $J_{up} \le 9$) associated with low-mass and intermediate-mass protostars \citep{2012A&A...542A..86Y, 2015A&A...581A...4L}, and the temperature of $\sim$50 K is slightly lower than the temperature for outflows of low-mass protostars \citep{2009A&A...501..633V, 2012A&A...542A..86Y} and intermediate-mass protostars \citep{2016A&A...587A..17V}. In addition, the derived outflow temperature is warmer than the previously adopted $\sim$30 K \citep{2009ApJ...696...66Q}, which indicates that the physical parameters (mass, momentum, energy) of the G240 outflow calculated by \citet{2009ApJ...696...66Q} were underestimated by a factor of 1.32. 

\subsection{Density-velocity relation}\label{subsec:density}
From the LVG analysis, the beam-averaged CO column density is well constrained in each velocity bin, while the constraint for the gas density is loose. Figure \ref{fig:figrelation}(b) shows that, from $\pm$7 km s$^{-1}$ to $\pm$22 km s$^{-1}$ with respect to the cloud velocity, the beam-averaged CO column density decreases by a factor of 50. The decreasing $N$-$V$ trend is similar to that of the outflow cavity associated with Cepheus E-mm \citep{2015A&A...581A...4L}, and may directly indicate a decrease of the entrained gas with increasing outflow velocity. Since the CO column density degenerates with the beam filling factor in the optically thin case, the observed CO column density could be affected by the beam dilution effect. However, the beam filling factor only decreases by a factor of $\sim$2.5 (derived from source sizes of $\sim$20$\arcsec$ to $\sim$10$\arcsec$, see Figure 3 of \citet{2009ApJ...696...66Q}) within the outflow velocity range, which could not fully explain the decrease of the observed CO column density with velocity. We argue that the decrease in CO column density with outflow velocity is due to the change of the gas density. Assuming a constant [CO]/[H$_2$] abundance ratio, which is adopted by most previous works, the decreasing CO column density indicates a decline of the gas column density with velocity. If the velocity gradient do not vary much, the decreasing gas column density with velocity implies that the gas density decreases with velocity. Taken together, the variations of the beam-averaged CO column density and the beam filling factor suggest that the gas density is $>$10$^6$ cm$^{-3}$ at the lowest outflow velocity. This gas density limit is comparable to that in the outflow cavity associated with Cepheus E-mm \citep[Several times of 10$^5$ cm$^{-3}$:][]{2015A&A...581A...4L}. In summary, our observations indicate that both the CO column density and the gas denstity decrease with an increasing outflow velocity.

\subsection{The origin of the G240 outflow}

With high-resolution CO J = 2--1 observations, \citet{2009ApJ...696...66Q} found that the G240 outflow has a well-defined, bipolar, wide-angle, quasi-parabolic morphology and a parabolic position-velocity structure in the northwest redshifted lobe. The kinematic structure and morphology are similar to low-mass wide-angle outflows. However, observations toward low-mass YSOs show that wide-angle low-velocity molecular outflows are usually accompanying with collimated high-velocity atomic/molecular jets: e.g., IRAS 04166+2706 \citep{2009A&A...495..169S}, L1448-mm \citep{2010ApJ...717...58H}, and HH 46/47 \citep{2007ApJ...668L.159V}. But the G240 region shows no signature of a high-velocity jet in infrared, millimeter, and centimeter emissions \citep{2002ApJ...576..313K, 2009ApJ...696...66Q, 2011AJ....142..147T}. Thus, it is still unclear whether the G240 outflow is driven in a way analogous to that of low-mass outflows or in a different way.

To further investigate the physical conditions and to explore the driven mechanism of the G240 outflow, we presented multi-line analysis of the outflow, and revealed that the temperature of the outflowing gas is relatively constant with outflow velocity, and that the density of the outflow decreases with gas velocity. These trends are in agreement with the estimations of a simple wide-angle wind-driven model, which predict that the outflow is approximately isothermal because of efficient cooling, and that the outflow density decreases with velocity and distance from the driving source as the wind sweeps up the ambient material, which has a density inversely proportional to the square of distance from the source \citep{1991ApJ...370L..31S, 2001ApJ...557..429L}. In contrast, the jet-driven bow shock model predicts that the gas temperature and the density of the outflow increase with velocity and distance from the driving source due to shock heating and compressing \citep{2001ApJ...557..429L}, which are different from our results. The none-detection of a high-velocity jet in the G240 region also rules out the jet-driven bow shock model. Although most existing wind-driven outflow models can only explain typical parameters of outflows driven by low-mass YSOs, recent theoretical results provide evidence that massive outflows can also be driven by wide-angle disk winds \citep{2018MNRAS.475..391M}. Thus, our analyses results support the scenario that the G240 outflow is mainly driven/entrained by a wide-angle wind, which itself may resemble the accretion-driven wide-angle winds \citep[X-wind or disk winds:][]{2006ApJ...649..845S, 2006MNRAS.365.1131P} associated with low-mass YSOs. We further suggest that disk-mediated accretion may exist in the formation of high-mass stars up to late-O types. 

\section{Summary}\label{summary}
Using the APEX CO J = 3--2, 6--5 and 7--6 observations and the complementary CO J = 2--1 data, we have presented the first CO multi-transition study of the molecular outflow in high-mass star-forming region \objectname{G240}. The parsec-sized, bipolar, and high velocity outflow is clearly detected in all the CO lines. For both blueshfited and redshifted lobes, the outflow is approximately isothermal with a temperature of $\sim$50 K. The CO column density of the outflow decreases with gas velocity, which indicates that the gas density decreases with outflow velocity if the CO abundance and velocity gradient do not significant vary. The isothermal state and the decreasing gas density provide compelling evidence that the well shaped massive outflow in G240 is driven by a wide-angle wind (rather than by jet bow-shocks). This finding further suggests that disk-accretion may be responsible for the formation of high-mass stars more massive than early B-type stars.

\acknowledgments
We thank the APEX staff for carrying out the observations, and Mr. Yu Cheng for helpful discussions on RADEX modeling. J.L., K.Q.,  Y. C., and Y.W. acknowledge supports from National Natural Science Foundation of China (Grant Nos. 11473011 and 11590781). This research made use of APLpy, an open-source plotting package for Python \citep{2012ascl.soft08017R}, Astropy, a community-developed core Python package for Astronomy \citep{2013A&A...558A..33A}, and Matplotlib, a Python 2D plotting library for Python \citep{2007CSE.....9...90H}.

\facility{Atacama Pathfinder Experiment (APEX).}

\software{APLpy \citep{2012ascl.soft08017R}, Astropy \citep{2013A&A...558A..33A},
          Matplotlib \citep{2007CSE.....9...90H}.}

\appendix
\section{Spectral line flux distributions}
\begin{figure*}[htbp]
\figurenum{A1}
\addtocounter{figure}{1}
\includegraphics[scale=.60]{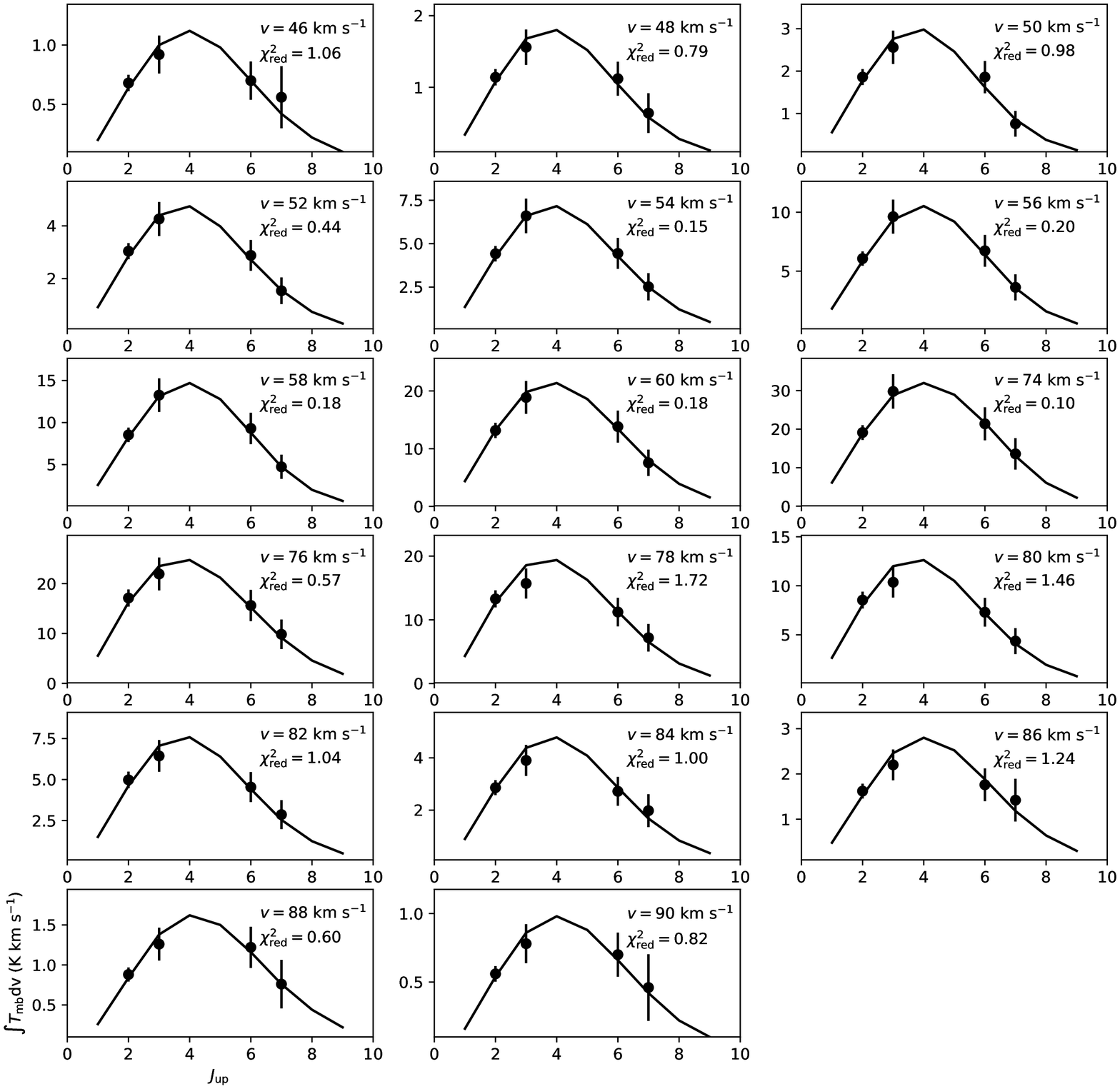}
\caption{Observed line fluxes compared with the LVG computations in each 2 km s$^{-1}$ bin. The black solid circles show the observed data with error bars. The black solid lines refer to the best fits.  The $\chi^2_{\mathrm{red}}$ of the best fitting results and the outflow velocities are shown in each panel. \label{fig:figsed}}
\end{figure*}

\end{CJK*}


\begin{thebibliography}{}
\bibitem[Arce et al.(2007)]{2007prpl.conf..245A} Arce, H.~G., Shepherd, D., Gueth, F., et al.\ 2007, Protostars and Planets V, 245 
\bibitem[Astropy Collaboration et al.(2013)]{2013A&A...558A..33A} Astropy Collaboration, Robitaille, T.~P., Tollerud, E.~J., et al.\ 2013, \aap, 558, A33
\bibitem[Bally (2016)]{Bally16}Bally, J. 2016, \araa, 54, 491
\bibitem[Banerjee \& Pudritz(2006)]{2006ApJ...641..949B} Banerjee, R., \& Pudritz, R.~E.\ 2006, \apj, 641, 949 
\bibitem[Beuther et al.(2002a)]{2002A&A...383..892B} Beuther, H., Schilke, P., Sridharan, T.~K., et al.\ 2002, \aap, 383, 892
\bibitem[Beuther et al.(2002b)]{2002A&A...387..931B} Beuther, H., Schilke, P., Gueth, F., et al.\ 2002, \aap, 387, 931 
\bibitem[Caswell(1997)]{1997MNRAS.289..203C} Caswell, J.~L.\ 1997, \mnras, 289, 203 
\bibitem[Caswell(2003)]{2003MNRAS.341..551C} Caswell, J.~L.\ 2003, \mnras, 341, 551 
\bibitem[Choi et al.(2014)]{2014ApJ...790...99C} Choi, Y.~K., Hachisuka, K., Reid, M.~J., et al.\ 2014, \apj, 790, 99 
\bibitem[Frank et al.(2014)]{2014prpl.conf..451F} Frank, A., Ray, T.~P., Cabrit, S., et al.\ 2014, Protostars and Planets VI, 451
\bibitem[Goldsmith \& Langer(1999)]{1999ApJ...517..209G} Goldsmith, P.~F., \& Langer, W.~D.\ 1999, \apj, 517, 209
\bibitem[Gomez-Ruiz et al.(2013)]{2013A&A...555A...8G} Gomez-Ruiz, A.~I., Wyrowski, F., Gusdorf, A., et al.\ 2013, \aap, 555, A8
\bibitem[Heyminck et al.(2006)]{2006A&A...454L..21H} Heyminck, S., Kasemann, C., G{\"u}sten, R., de Lange, G., \& Graf, U.~U.\ 2006, \aap, 454, L21 
\bibitem[Hirano et al.(2010)]{2010ApJ...717...58H} Hirano, N., Ho, P.~P.~T., Liu, S.-Y., et al.\ 2010, \apj, 717, 58
\bibitem[Hughes \& MacLeod(1993)]{1993AJ....105.1495H} Hughes, V.~A., \& MacLeod, G.~C.\ 1993, \aj, 105, 1495
\bibitem[Hunter (2007)]{2007CSE.....9...90H} Hunter, J.~D.\ 2007, Computing in Science and Engineering, 9, 90 
\bibitem[Kasemann et al.(2006)]{2006SPIE.6275E..0NK} Kasemann, C., G{\"u}sten, R., Heyminck, S., et al.\ 2006, \procspie, 6275, 62750N 
\bibitem[Kumar et al.(2002)]{2002ApJ...576..313K} Kumar, M.~S.~N., Bachiller, R., \& Davis, C.~J.\ 2002, \apj, 576, 313 
\bibitem[Kumar et al.(2003)]{2003A&A...412..175K} Kumar, M.~S.~N., Fernandes, A.~J.~L., Hunter, T.~R., Davis, C.~J., \& Kurtz, S.\ 2003, \aap, 412, 175 
\bibitem[Lee et al.(2000)]{2000ApJ...542..925L} Lee, C.-F., Mundy, L.~G., Reipurth, B., Ostriker, E.~C., \& Stone, J.~M.\ 2000, \apj, 542, 925
\bibitem[Lee et al.(2001)]{2001ApJ...557..429L} Lee, C.-F., Stone, J.~M., Ostriker, E.~C., \& Mundy, L.~G.\ 2001, \apj, 557, 429 
\bibitem[Lee et al.(2002)]{2002ApJ...576..294L} Lee, C.-F., Mundy, L.~G., Stone, J.~M., \& Ostriker, E.~C.\ 2002, \apj, 576, 294 
\bibitem[Lefloch et al.(2015)]{2015A&A...581A...4L} Lefloch, B., Gusdorf, A., Codella, C., et al.\ 2015, \aap, 581, A4
\bibitem[Li et al.(2013)]{2013A&A...559A..23L} Li, G.-X., Qiu, K., Wyrowski, F., \& Menten, K.\ 2013, \aap, 559, A23 
\bibitem[Machida et al.(2008)]{2008ApJ...676.1088M} Machida, M.~N., Inutsuka, S.-i., \& Matsumoto, T.\ 2008, \apj, 676, 1088-1108  
\bibitem[MacLeod et al.(1998)]{1998AJ....116.1897M} MacLeod, G.~C., Scalise, E., Jr., Saedt, S., Galt, J.~A., \& Gaylard, M.~J.\ 1998, \aj, 116, 1897 
\bibitem[Masson \& Chernin(1993)]{1993ApJ...414..230M} Masson, C.~R., \& Chernin, L.~M.\ 1993, \apj, 414, 230 
\bibitem[Matsushita et al.(2018)]{2018MNRAS.475..391M} Matsushita, Y., Sakurai, Y., Hosokawa, T., \& Machida, M.~N.\ 2018, \mnras, 475, 391 
\bibitem[Maud et al.(2015)]{2015MNRAS.453..645M} Maud, L.~T., Moore, T.~J.~T., Lumsden, S.~L., et al.\ 2015, \mnras, 453, 645
\bibitem[Migenes et al.(1999)]{1999ApJS..123..487M} Migenes, V., Horiuchi, S., Slysh, V.~I., et al.\ 1999, \apjs, 123, 487
\bibitem[Pudritz et al.(2006)]{2006MNRAS.365.1131P} Pudritz, R.~E., Rogers, C.~S., \& Ouyed, R.\ 2006, \mnras, 365, 1131 
\bibitem[Pudritz et al.(2007)]{2007prpl.conf..277P} Pudritz, R.~E., Ouyed, R., Fendt, C., \& Brandenburg, A.\ 2007, Protostars and Planets V, 277
\bibitem[Qiu et al.(2009)]{2009ApJ...696...66Q} Qiu, K., Zhang, Q., Wu, J., \& Chen, H.-R.\ 2009, \apj, 696, 66
\bibitem[Qiu et al.(2014)]{2014ApJ...794L..18Q} Qiu, K., Zhang, Q., Menten, K.~M., et al.\ 2014, \apjl, 794, L18 
\bibitem[Raga \& Cabrit(1993)]{1993A&A...278..267R} Raga, A., \& Cabrit, S.\ 1993, \aap, 278, 267 
\bibitem[Ren et al.(2011)]{2011MNRAS.415L..49R} Ren, J.~Z., Liu, T., Wu, Y., \& Li, L.\ 2011, \mnras, 415, L49 
\bibitem[Robitaille \& Bressert(2012)]{2012ascl.soft08017R} Robitaille, T., \& Bressert, E.\ 2012, Astrophysics Source Code Library, ascl:1208.017 
\bibitem[Sakai et al.(2015)]{2015PASJ...67...69S} Sakai, N., Nakanishi, H., Matsuo, M., et al.\ 2015, \pasj, 67, 69
\bibitem[Santiago-Garc{\'{\i}}a et al.(2009)]{2009A&A...495..169S} Santiago-Garc{\'{\i}}a, J., Tafalla, M., Johnstone, D., \& Bachiller, R.\ 2009, \aap, 495, 169
\bibitem[Shang et al.(2006)]{2006ApJ...649..845S} Shang, H., Allen, A., Li, Z.-Y., et al.\ 2006, \apj, 649, 845 
\bibitem[Shepherd et al.(1998)]{1998ApJ...507..861S} Shepherd, D.~S., Watson, A.~M., Sargent, A.~I., \& Churchwell, E.\ 1998, \apj, 507, 861 
\bibitem[Shu et al.(1991)]{1991ApJ...370L..31S} Shu, F.~H., Ruden, S.~P., Lada, C.~J., \& Lizano, S.\ 1991, \apjl, 370, L31
\bibitem[Shu et al.(2000)]{2000prpl.conf..789S} Shu, F.~H., Najita, J.~R., Shang, H., \& Li, Z.-Y.\ 2000, Protostars and Planets IV, 789
\bibitem[Su et al.(2012)]{2012ApJ...744L..26S} Su, Y.-N., Liu, S.-Y., Chen, H.-R., \& Tang, Y.-W.\ 2012, \apjl, 744, L26 
\bibitem[Trinidad(2011)]{2011AJ....142..147T} Trinidad, M.~A.\ 2011, \aj, 142, 147 
\bibitem[van der Tak et al.(2007)]{2007A&A...468..627V} van der Tak, F.~F.~S., Black, J.~H., Sch{\"o}ier, F.~L., Jansen, D.~J., \& van Dishoeck, E.~F.\ 2007, \aap, 468, 627
\bibitem[van Kempen et al.(2009a)]{2009A&A...501..633V} van Kempen, T.~A., van Dishoeck, E.~F., G{\"u}sten, R., et al.\ 2009, \aap, 501, 633 
\bibitem[van Kempen et al.(2009b)]{2009A&A...507.1425V} van Kempen, T.~A., van Dishoeck, E.~F., G{\"u}sten, R., et al.\ 2009, \aap, 507, 1425 
\bibitem[van Kempen et al.(2016)]{2016A&A...587A..17V} van Kempen, T.~A., Hogerheijde, M.~R., van Dishoeck, E.~F., et al.\ 2016, \aap, 587, A17
\bibitem[Velusamy et al.(2007)]{2007ApJ...668L.159V} Velusamy, T., Langer, W.~D., \& Marsh, K.~A.\ 2007, \apjl, 668, L159 
\bibitem[Wu et al.(2004)]{2004A&A...426..503W} Wu, Y., Wei, Y., Zhao, M., et al.\ 2004, \aap, 426, 503
\bibitem[Wu et al.(2005)]{2005AJ....129..330W} Wu, Y., Zhang, Q., Chen, H., et al.\ 2005, \aj, 129, 330 
\bibitem[Xie \& Qiu(2018)]{2018RAA....18...19X} Xie, Z.-Q., \& Qiu, K.-P.\ 2018, Research in Astronomy and Astrophysics, 18, 019
\bibitem[Y{\i}ld{\i}z et al.(2012)]{2012A&A...542A..86Y} Y{\i}ld{\i}z, U.~A., Kristensen, L.~E., van Dishoeck, E.~F., et al.\ 2012, \aap, 542, A86
\bibitem[Zhang et al.(2001)]{2001ApJ...552L.167Z} Zhang, Q., Hunter, T.~R., Brand, J., et al.\ 2001, \apjl, 552, L167 
\end{thebibliography}
\end{document}